\documentclass[aps,prl,twocolumn,showpacs,preprintnumbers,amsmath,amssymb,floatfix]{revtex4}

\usepackage{graphicx}

\begin{document}

\title{The von Neumann-Wigner theorem in quantum dot molecules}

\author{L.-X. Zhang, D. V. Melnikov, and J.-P. Leburton}
\affiliation{
Beckman Institute for Advanced Science \& Technology and Department of Electrical and Computer Engineering,
University of Illinois at Urbana-Champaign, Urbana, Illinois 61801
}

\date{\today}
\begin{flushleft}

\end{flushleft}
\begin{abstract}

We show that electrons in coupled quantum dots characterized by high aspect ratios undergo abrupt density rotations when the dots are biased into an asymmetric confinement configuration. Density rotations occur with electron transfer to a single dot, and give rise to sharp variations of the exchange coupling between electrons as a function of inter-dot detuning. Our analysis based on exact diagonalization technique indicates that this unusual behavior is in agreement with the von Neumann-Wigner theorem that dictates the variations of the energy spectrum from the symmetries of the molecular states during the detuning process. It is also shown that the overall effect is quenched by the presence of magnetic fields, which by adding angular momentum to the electron motion affects the spatial symmetry of the molecular states.

\end{abstract}


\maketitle

The experimental observation of atomic-like properties in manmade semiconductor quantum dots (QD) with feature sizes of a few hundreds to thousands of Angstroms has stimulated intense research in the last decade \cite{Ashoori, Kastner, Gordon,Ciorga,Reimann, Wiel}. The existence of shell structures and the demonstration of Hund's first rule with shell filling controlled by electric gates in QDs have provided the conditions for investigating tunable many-body effects in solid state systems \cite{Tarucha,Matagne}. As natural extensions of the concept of artificial atoms, coupled QDs resemble natural molecules and offer enhanced many-body tunability by electrically varying the coupling barrier or distance between the dots \cite{Waugh,Elzerman}, and detuning the energy spectra of the individual QDs by appropriate gate biasing \cite{Petta1}, which are fundamental technical ingredients for realizing a quantum logic gate \cite{Loss,Hanson}

In this context, the symmetry of the electron wave functions is of crucial importance for the evolution of the energy spectrum of molecular systems under the influence of perturbations. According to the von Neumann-Wigner theorem \cite{Wigner}, energy levels cross for states that bear different symmetries, while they anti-cross for states with the same symmetry. This fundamental theorem has found successful applications in the spectroscopy of alkaline salts \cite{Bron}, the interpretation of the Zeeman spectrum of hydrogen molecules \cite{Lichten}, and more recently in the explanation of the magnetic anisotropy in thin metallic films \cite{Pick}.
       
Spatial symmetry effects can be mostly evidenced in double QDs with large aspect ratios, {\it i.e.}, for which the confinement strength in the coupling direction is comparable to or larger than that in the perpendicular direction. Hence inter-dot detuning is expected to induce significant conformational changes in the electron wave functions, which rearranges the electronic spectra of the individual QDs with drastic effects on the many-body interaction. 

\begin{figure}[b]
\includegraphics[width=6.9cm]{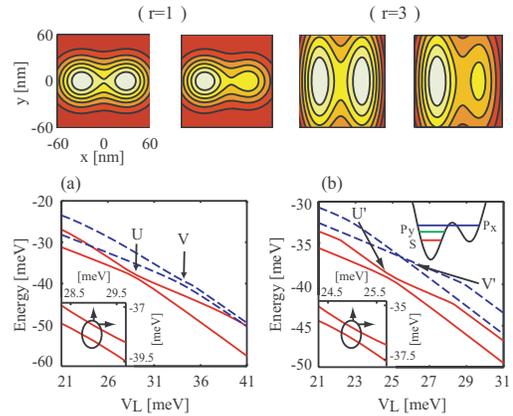}
\caption{\label{fig:fig1} (color online) Top panels: potential contour plots of coupled QD with $r=1$ (left two) and $r=3$ (right two). For each $r$, the panel on the left (right) is for $V_L=V_R=21$ ($V_L=31$, $V_R=21$) meV. (a) $r=1$. Main panel: two lowest singlet (red, solid lines) and triplet (blue, dashed lines) energy levels as a function of detuning. $U$ ($V$) indicates anti-crossing points for the two lowest singlet (triplet) levels. Lower inset: zoom-in region near the anti-crossing point of the singlet energy levels. (b) $r=3$. Main panel and lowest inset same as in (a) but for $r=3$. $U^\prime$ ($V^\prime$) indicates the anti-crossing (crossing) point of the two lowest singlet (triplet) energy levels \cite{footnote2}. Upper inset: three lowest single-particle energy levels in asymmetric confinement. In (a) and (b), the singlet energy levels are lowered by $3$ meV for clarity.}
\end{figure}
                     
In this Letter, we show that two electrons in coupled QDs with non-zero ellipticity experience density rotation and abrupt change of the ordering of their energy levels with inter-dot detuning. The charge density rotation occurs at specific detuning values, and leads to a discontinuity in the derivative of the exchange energy $J$ with respect to the detuning bias, which is in agreement with the von Neumann-Wigner theorem \cite{Wigner}. We also show that the presence of magnetic fields enhances electron localization in the QDs and breaks the spatial symmetry in the two-electron states, which quenches the density rotation effect.

\begin{figure}[t]
\includegraphics[width=6.8cm]{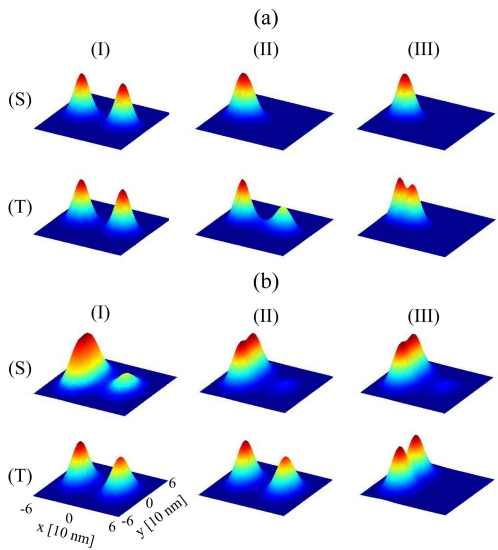}
\caption{\label{fig:fig2} (color online) (a) Two-dimensional density plots for $r=1$ at zero magnetic field. Columns (I), (II) and (III) correspond to $V_L=21$, $29$ and $42$ meV, respectively. First (second) row, labeled S (T), is for the lowest singlet (triplet) states.  (b) Same as (a) but for $r=3$. Columns (I), (II) and (III) correspond to $V_L=25$, $25.45$ and $25.47$ meV, respectively. Coordinates are shown in lower left panel and are the same for all panels. For $r=3$, the peak separation along the $x$ ($y$) direction in the density of the triplet state is $\sim 60$ ($\sim 40$) nm for column (II) [(III)].}
\end{figure}

We model the coupled QDs with the {\it ad hoc} potential $ V({\bf r})=-V_Lexp[-(x+d/2)^2/R_x^2-y^2/R_y^2]-V_Rexp[-(x-d/2)^2/R_x^2 -y^2/R_y^2]$, where $d$ defines the inter-dot separation, $R_x$ and $R_y$ are the extensions of the Gaussian potential for each dot in the coupling direction $x$ and perpendicular direction $y$. In this work, we take $d=60$ nm, $R_x=30$ nm, and define the QD aspect ratio as $r = R_y/R_x$. The difference between $V_L$ and $V_R$ gives the inter-dot detuning $\epsilon = V_L-V_R$. We fix $V_R$ at $21$ meV and increase $V_L$ from $21$ meV to higher values, which is equivalent to increasing $\epsilon$. Figure \ref{fig:fig1}, top panels show the potential contour plots for circular (r=1, left two panels) and elliptical (r=3, right two panels) QDs under symmetric and asymmetric biases. As $\epsilon$ increases the coupled QD system becomes effectively a single QD centered at the lowest point of the confinement potential (the left dot center for both cases).

We use numerical exact diagonalization of the two-electron Hamiltonian to solve for the system energies within the effective mass approximation. Details for the method and the material parameters are published elsewhere \cite{Zhang2}. The convergence of the energies is tested by comparing the results using $8$ and $9$ harmonic oscillator basis states in each Cartesian direction (we use $9$ states in each direction for this work). We ignore the Zeeman effect which simply lowers the triplet energy and $J$ by $\sim 25$ ${\mu}$eV/T at a given magnetic field. 
  
\begin{figure}[t]
\includegraphics[width=6.5cm]{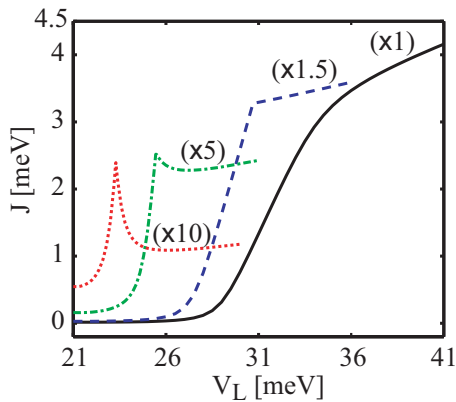}
\caption{\label{fig:fig3} (color online) Exchange coupling $J$ as a function of $V_L$ ($V_R$ fixed at $21$ meV) for QD aspect ratios $r=1$ (black, solid), $r=1.5$ (blue, dashed), $r=3$ (green, dashed-dotted) and $r=5$ (red, dotted) at zero magnetic field. The scaling factor for each curve is shown in parenthesis.} 
\end{figure}

In Fig. \ref{fig:fig1}, bottom panels, we plot the two lowest singlet and the two lowest triplet energy levels for the coupled QDs with $r=1$ [Fig. \ref{fig:fig1}(a)] and $r=3$ [Fig. \ref{fig:fig1}(b)] as a function of detuning $\epsilon$ (from here on, if not specified, ``singlet'' and ``triplet'' refer to the singlet and triplet states of lowest energy, respectively). All energy levels decrease as $V_L$ increases because the single-particle energies that give the main contribution to the two-electron energy are lowered with the electric potential. For both aspect ratios, $r=1$ and $3$, and at small $\epsilon$ [$21$ $\le$ $V_L$ $\le$ $29$ meV in Fig. \ref{fig:fig1}(a) and $21$ $\le$ $V_L$ $\le$ $25$ meV Fig. \ref{fig:fig1}(b)], we observe singlet and triplet energies decrease with nearly the same slope as $\epsilon$ increases. Then, the singlet energy slope becomes steeper as both electrons localize into one QD, which are indicated by points $U$ and $U^\prime$ in Fig. \ref{fig:fig1}. For $r=1$ and $V_L>34$ meV, the slope of the triplet energy increases in value (yet smaller than that of singlet), while for $r=3$ the triplet energy experiences a {\it sudden} change of slope at $V_L=25.46$ meV [point $V^\prime$ in Fig. \ref{fig:fig1}(b)], and with increasing $V_L$ values afterwards, the spacing between singlet and triplet energies first decreases, and then gradually increases for $V_L \geq 27.2$ meV (cf. $J$ curve in Fig. \ref{fig:fig3}).

The detuning effects on the singlet and triplet states can be further illustrated by inspecting the electron density variations. In Fig. \ref{fig:fig2}(a) we see that for coupled circular QDs ($r=1$) with increasing $\epsilon$ electrons gradually move into the lower (left) QD for both singlet and triplet states, albeit more quickly for the singlet state. When both electrons are in the left dot [Fig. \ref{fig:fig2}(a), column (III)], the singlet density has a single maximum corresponding to the state with nearly zero angular momentum \cite{Wagner}, while  the triplet state exhibits two maxima along the QD coupling direction ($x$) as the constituent QDs are slightly more extended in the $x$-direction than in the $y$-direction. For $r=3$, as $\epsilon$ increases, the singlet density gradually localizes into the left dot with the electron density showing two maxima in the $y$-direction once $V_L>25$ meV because of the relaxed confinement in that direction [cf. Fig. \ref{fig:fig2}(b) columns (I) and (II)]. For the triplet state, however, it is seen that the electron density abruptly changes from being spread over the two QDs with a higher peak in the left dot [$V_L=25.45$ meV, column (II) in Fig. \ref{fig:fig2}(b)] to occupying only the left dot with two peaks of equal height in the $y$-direction [$V_L=25.47$ meV, column (III) in Fig. \ref{fig:fig2}(b)]. This abrupt transition (or ``rotation'') of the electron density occurs at $V_L=25.46$ meV [point $V^\prime$ in Fig. \ref{fig:fig2}(b)].

In order to understand the drastic difference in the dependences of the energy levels on $\epsilon$ between the QD configurations with $r=1$ and $r=3$, we compute the expectation values of the parity operator $\left\langle \hat{P} \right\rangle =\left\langle \Psi(x_1,y_1,x_2,y_2)| \Psi(-x_1,-y_1,-x_2,-y_2) \right\rangle$, and of the parity operator with respect to the $y$-axis $\left\langle \hat{P}_y \right\rangle=\left\langle \Psi(x_1,y_1,x_2,y_2)| \Psi(x_1,-y_1,x_2,-y_2) \right\rangle $. Here, $\Psi$ is the two-electron wavefunction, and ($x_1$, $y_1$), ($x_2$, $y_2$) are the coordinates of the first and second electrons, respectively. In the investigated detuning range, we find that for the singlet state, $0<P<1$ and $P_y=1$ for all $r$; while for the triplet state at $r=1$, $P<0$ and $P_y=1$, as expected from the symmetry of the Hamiltonian. However, in the triplet state for $r=3$ at $V_L=25.46$ meV, our calculation shows that the value of $\left\langle \hat{P}\right\rangle$ sharply increases from $-0.98$ to $-0.02$, while $\left\langle \hat{P}_y\right\rangle$ changes sign from $1$ to $-1$. This indicates that the parity of the singlet state with respect to the $y$-axis remains even, while the triplet wavefunction parity changes from even to odd, thereby changing its symmetry along the $y$-direction. In fact, we find that as the inter-dot detuning is increased, the two lowest singlet states at any $r$ have the same symmetry, and their energy levels anticross; the same is true for the two lowest triplet states at $r=1$ (see Fig. \ref{fig:fig1}), whereas the energy levels of the two lowest triplet states at $r>1$ [Fig. \ref{fig:fig1}(b)] cross due to their different symmetries in agreement with the von Neumann-Wigner theorem \cite{Wigner}. Near the crossing point of the two lowest triplet energy levels [point $V^\prime$ in Fig. \ref{fig:fig1}(b)], the analysis of the spectral function \cite{Dmm2} reveals that the first excited single-particle state localized in the left QD has the $p_y$-character [see schematic energy diagram in the upper inset of Fig. \ref{fig:fig1}(b)]. 
Before the level crossing, this $p_y$-like orbital is unoccupied so that the electronic states form a $sp_x$-pair consisting of the $s$-like state mostly localized in the left QD, and the lowest antisymmetric $p_x$-like state spreading over both QDs. After the level crossing, it becomes energetically favorable for the triplet electrons to be in the same QD as an $sp_y$-pair, even though the expectation value of the Coulomb interaction in the triplet state increases from 2.0 meV before the transition $sp_x\rightarrow sp_y$ to 2.9 meV afterwards.

Figure \ref{fig:fig3} displays the exchange coupling $J$ as a function of $V_L$, or effectively as a function of $\epsilon=V_L-V_R$. Each curve in Fig. \ref{fig:fig3} depicts the variation of $J$ as the coupled QDs undergo the transition from the $(1,1)$ to $(2,0)$ charge states (the right end of each curve is chosen at a bias point where $\epsilon$ is large enough to localize both electrons into the left dot). For small departure from equal biases ($\epsilon \gtrsim 0$), as $V_L$ increases from $21$ meV, $J$ maintains a small and approximately constant value before it increases superlinearly, as was recently observed both experimentally \cite{Petta1} and reproduced theoretically \cite{Stopa}. For $r=1$, $J$ continues to increase monotonically with $\epsilon$ as the two electrons strongly localize in one QD. In this case, the localization of the first excited single-particle state that has $p_x$-character in the left dot occurs at larger detuning value than the ground state [points $U$ and $V$ in Fig. \ref{fig:fig1}(a)], and  $J$ can only increase as the inter-level separation in the individual QD is larger than that in the coupled QD system. However, with increasing aspect ratios ($r \geq 1.5$), dramatic changes in the $J$ behavior appear: a kink in $J$ occurs at $V_L=30.70$ meV with a sudden slope change for $r=1.5$, while sharp maxima in $J$ arise at $25.46$ and $23.30$ meV for $r=3$ and $r=5$, respectively. In the latter cases, there is also a local $J$ minimum at $V_L=27.20$ and $26.20$ meV, respectively. For increasing $r$, electron localization in the left dot occurs at progressively smaller detuning (Fig. \ref{fig:fig3}), as it becomes easier for the system to lower its energy in a configuration where the two electrons are located at the opposite ``ends'' of the elliptical QD [Fig. \ref{fig:fig2}(b)] to minimize their Coulomb interaction. As a result, both singlet and triplet electrons move into the left dot at close $V_L$ values  [{\it e.g.}, points $U^\prime$ and $V^\prime$ in Fig. \ref{fig:fig1}(b) for $r=3$]. Depending upon the difference between the biases required to localize singlet and triplet states into a single dot, the crossing between the two lowest triplet states associated with the electron density rotation is observed as a kink ($r\gtrsim 1$) or a sharp maximum ($r\gg 1$) in the exchange coupling at intermediate values of detuning. 

\begin{figure}[t]
\includegraphics[width=6.8cm]{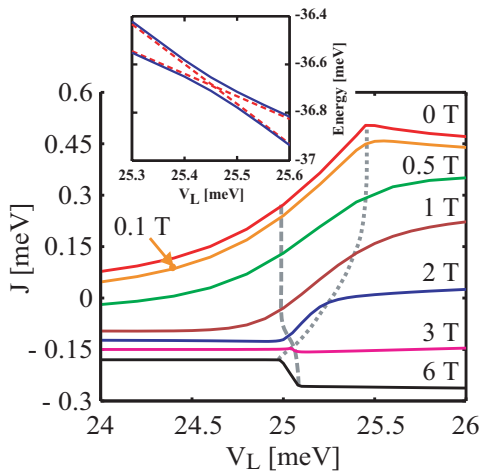}
\caption{\label{fig:fig4} (color online) Main panel: $J$ as a function of $V_L$ for  $r=3$ at different magnetic fields. The magnetic field value is given on top of each curve. The curve for $B=0.1$ T is indicated by an arrow. For clarity, curves at different $B$ fields are shifted vertically in multiples of $0.03$ meV. The dashed (dotted) gray line is a guide for the eyes tracing the localization point of the singlet (triplet) state at different magnetic fields. Inset: two lowest triplet energy levels at $B=0$ T (red, dashed) and $B = 0.1$ T (blue, solid).} 
\end{figure}

Figure \ref{fig:fig4}, main panel shows the $J$-dependence on $V_L$ for $r=3$ in the detuning range corresponding to the $(1,1) \Rightarrow (2,0)$ transition ($24 \leq V_L \leq 26 $ meV; $V_R=21$ meV) for different magnetic fields normal to the $xy$-plane. We observe that the sharp discontinuities in $dJ/dV_L$ due to abrupt electron density rotation with detuning disappears for $0<B \leq 2$ T, and evolves into a smooth $J$-increase with $V_L$. By adding angular momentum to the electron motion, the magnetic field mixes the $p_x$ and $p_y$ state parities, which affect the symmetry of the two-electron wavefunction, and quenches the abruptness of the density rotation. This symmetry breaking (from $p_y$-symmetry to $p_x$-$p_y$ symmetry mixing) removes the crossing conditions between the two lowest triplet states to produce level anti-crossing, in agreement with the von Neumann-Wigner theorem (Fig. \ref{fig:fig4} inset). Our analysis also indicates smooth variations of the electron density with detuning at small non-zero ( $0<B \leq 2$ T) magnetic fields: for small $\epsilon$, the triplet electron density has a higher peak in the left dot than in the right dot, which, with increasing $\epsilon$, evolves into two peaks along the $y$-direction, while the peak in the right dot gradually disappears. At higher magnetic fields ($B=3$ and $6$ T),  the localization of both singlet and triplet states in the left dot occurs abruptly with increasing detuning because of the strong magnetic compression of the electron orbitals. We observe that for increasing magnetic fields the triplet state localization requires progressively smaller detuning [shown by the dotted line on Fig. \ref{fig:fig4}(b)], while the singlet state localization occurs at almost the same detuning [shown by the dashed line on Fig. \ref{fig:fig4}(b)] for all investigated magnetic fields \cite{footnote3}.

Experimentally, the predicted variations in the two-electron density and exchange coupling can be obtained from the conductance responses of quantum point contacts adjacent to the dots, which are sensitive enough to register very slight changes in the Coulomb potential distributions in the coupled QD system \cite{Elzerman,Petta1}. One can also detect the existence of the maximum in the exchange coupling from the coherent Rabi oscillations in the two-electron system by properly biasing the QD system \cite{Petta1}. Finally, electron imaging spectroscopy can be utilized to directly probe the electron density in QDs \cite{Fallahi}.

In conclusion, we show that inter-dot detuning in coupled elliptical quantum dots induces abrupt density rotations. We find that at zero magnetic field as the detuning increases, energy level crossing occurs for the two lowest triplet states bearing different (even/odd) symmetry. Depending on the dot aspect ratio $r$, such a crossing results in a kink ($r\gtrsim 1$) or sharp local maximum ($r\gg 1$) in the dependence of the exchange coupling on the detuning. For small nonzero magnetic field, the triplet energy level crossing in coupled dots with large aspect ratios evolves into anti-crossing due to the mixing of single-particle states bearing different spatial symmetry. For even larger magnetic fields, the localization of electrons by detuning is more abrupt due to the strong compression of electronic orbitals. 

This work is supported by the DARPA QUIST program and NSF through the Material Computational Center at the University of Illinois. LXZ thanks the Beckman Institute and Computer Science and Engineering program at the University of Illinois.


\begin{thebibliography}{99}

\bibitem {Ashoori} R. C. Ashoori, Phys. Rev. Lett. {\bf 68}, 3088 (1992).

\bibitem {Kastner} M. A. Kastner, Physics Today {\bf 46}, 24 (1993).

\bibitem {Gordon} D. Goldhaber-Gordon {\it et al.}, Nature {\bf 391}, 156 (1998).

\bibitem {Ciorga} M. Ciorga {\it et al.},  Phys. Rev. B {\bf 61}, R16315 (2000).

\bibitem{Reimann} S. M. Reimann and M. Manninen, Rev. Mod. Phys. {\bf 74}, 1283 (2002).

\bibitem{Wiel} W. G. van der Wiel {\it et al.}, Rev. Mod. Phys. {\bf 75}, 1 (2003).

\bibitem {Tarucha} S. Tarucha {\it et al.}, Phys. Rev. Lett. {\bf 77}, 3613 (1996).

\bibitem {Matagne} P. Matagne {\it et al.}, Phys. Rev. B {\bf 65}, 085325 (2002).

\bibitem {Waugh} F. R. Waugh {\it et al.}, Phys. Rev. Lett. {\bf 75}, 705 (1995).

\bibitem {Elzerman} J. M. Elzerman {\it et al.}, Phys. Rev. B {\bf 67}, 161308(R) (2003);

\bibitem{Petta1} J. R. Petta {\it et al.}, Science {\bf 309}, 2180 (2005);

\bibitem{Loss} D. Loss and D. P. DiVincenzo, Phys. Rev. A {\bf 57}, 120 (1998).

\bibitem{Hanson} R. Hanson {\it et al.},  Rev. Mod. Phys. {\bf 79}, 1217 (2007).

\bibitem{Wigner} J. von Neumann and E. Wigner,  Z. Phys. {\bf 30}, 467 (1929); L. D. Landau and E. Lifshitz, {\it Quantum Mechanics: Non-Relativistic Theory} (Pergamon Press, Oxford, 1977).

\bibitem{Bron} W.E. Bron and M. Wagner, Phys. Rev. {\bf 145}, 689 (1966).

\bibitem{Lichten} W. Lichten, Phys. Rev. A {\bf 3}, 594 (1971).

\bibitem{Pick} \v{S}, Pick and H. Dreyss\'{e}, Phys. Rev. B {\bf 48}, 13588 (1993).

\bibitem{Zhang2} L.-X. Zhang {\it et al.}, Phys. Rev. B {\bf 74}, 205306 (2006).

\bibitem{footnote2} At $V_L=25.46$ meV, the computed spacing between the two lowest triplet energy levels is less than $\sim1$~$\mu$eV. By the inspection of the energy level slopes and wavefunction symmetry, we conclude that a crossing occurs at this point.

\bibitem{Wagner} M. Wagner {\it et al.}, Phys. Rev. B {\bf 45}, R1951 (1992). 

\bibitem{Dmm2} D. V. Melnikov and J.-P. Leburton, Phys. Rev. B {\bf 73}, 155301 (2006).

\bibitem{Stopa} M. Stopa and C. M. Marcus, cond-mat/0604008 (2006).

\bibitem{footnote3} The localization of the singlet and triplet states is determined at the point where the two lowest two-particle
energy levels cross or anti-cross as detuning increases.

\bibitem{Fallahi} P. Fallahi {\it et al.}, Nano Lett. {\bf 5}, 223 (2005).

\end{thebibliography}
\end{document}